\documentclass[final,5p,times,twocolumn]{elsarticle} 

\usepackage{lineno}
\usepackage{siunitx}
\usepackage{graphicx}
\usepackage{subcaption}
\usepackage{multirow}
\usepackage{lineno,hyperref}
\modulolinenumbers[5]

\journal{VCI Proceedings in NIM A}

\begin{document}

\begin{frontmatter}

\title{Irradiation Studies of the Resistive AC-coupled Silicon Detector (RSD/AC-LGAD)}

\author[a]{Umut Elicabuk,}
\author[a]{Brendan Regnery\corref{cor1}}
\author[b,c]{Luca Menzio,}
\author[b,c]{Roberta Arcidiacono,}
\author[b]{Nicolo Cartiglia,}
\author[a]{Alexander Dierlamm,}
\author[a]{Markus Klute,}
\author[b]{Marco Ferrero,}
\author[a]{Ling Leander Grimm,}
\author[d,e]{Francesco Moscatelli,}
\author[b]{Federico Siviero}
\author[f]{Matteo Centis Vignali}

\cortext[cor1]{corresponding author\\ \textit{email:}~brendan.regnery@kit.edu}

\address[a]{Karlsruhe Institute of Technology, Karlsruhe, Germany}
\address[b]{Sezione di Torino, Istituto Nazionale di Fisica Nucleare, Via Pietro Giuria 1, 10125 Torino, Italy}
\address[c]{Dipartimento di Scienze del Farmaco, Universita del Piemonte Orientale, Largo Donegani 2, 28100 Novara, Italy}
\address[d]{INFN Sezione di Perugia, Perugia, Italy}
\address[e]{Consiglio Nazionale delle Ricerche - Istituto Officina dei Materiali, Perugia, Italy}
\address[f]{Fondazione Bruno Kessler, Trento, Italy}

\begin{abstract}
Resistive AC-coupled Silicon Detectors (RSDs) are silicon sensors which provide high temporal and spatial resolution. The RSD is a candidate sensor to be used in future tracking detectors with the objective of obtaining '4D' tracking where timing information can be used along with spatial hits during track finding. 4D tracking will be an essential part of any future lepton or hadron collider and may even be feasible at the HL-LHC. For applications at hadron colliders, RSD sensors must be able to operate in high fluence environments in order to provide 4D tracking. However, the effects of radiation on RSDs have not been extensively studied. In this study, RSDs were irradiated to 1.0, 2.0, $3.5\times10^{15}$~\si{\centi\meter^{-2}} 1~\si{\mega\electronvolt} neutron equivalences with both protons and neutrons. The sensors were then characterized electrically to study the acceptor removal and, for the first time in this doping concentration range, the donor removal. Then, the Transient Current Technique was used to begin investigating the signal charge sharing after irradiation. The results suggest an interesting trend between acceptor and donor removal, which is worthy of further study and could assist in improving radiation hardness of Low Gain Avalanche Diodes (LGADs).
\end{abstract}


\end{frontmatter}


\section{Introduction}

The past decade of developments in Low Gain Avalanche Diodes (LGADs) has culminated in $1.3\times1.3$~\si{\milli\meter^{2}} sensors capable of $<50$~\si{\pico\second} time resolutions in the high radiation environments expected at the HL-LHC \cite{ATLAS_HGTD, CMS_MTD, FrankBook}. These sensors are now being used in the phase-2 upgrade projects at CMS and ATLAS for timing layers. The evolution of LGADs from simple R\&D projects to full scale production detectors has motivated an entirely new generation of detectors capable of high spatial and temporal resolution. 

As with timing layers, constructing a tracking layer with such sensors can further improve pile-up discrimination in high rate environments (e.g. HL-LHC, FCC-hh) or provide time-of-flight information used for particle identification (e.g. FCC-ee). Additionally, constructing a full tracking system with multiple layers of such sensors allows for the possibility of '4D' tracking. '4D' tracking would allow for the inclusion of timing information while tracks are created, thus greatly reducing track finding combinatorics.

\begin{figure}[htbp]
    \centering
    \includegraphics[width=0.65\linewidth]{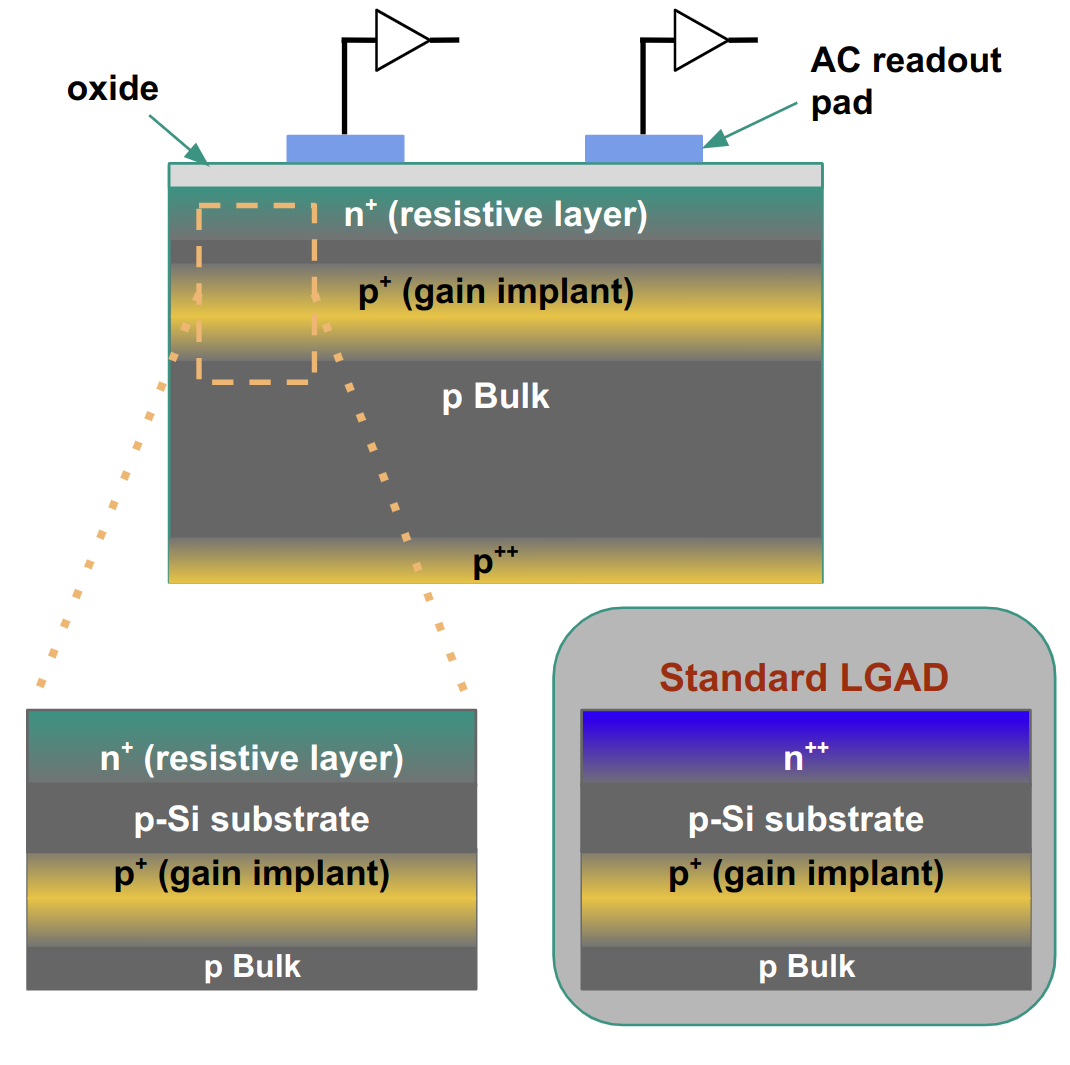}
    \caption{An RSD utilizes charge sharing on AC pads to provide high spatial resolution. A key difference between the RSD and a standard LGAD is the n+ resistive layer which is significantly less doped than the n++ layer in a standard LGAD in order to provide the high resistivity needed to allow for charge sharing.}
    \label{fig:RSD-LGAD-comp}
\end{figure}

Several novel ideas for devices to enable '4D' tracking are under study, one such idea is the Resistive AC-coupled Silicon Detector (RSD)~\cite{NicoloBook, RSD1MarcoM, RSD1Marta}. The RSD sensors utilize a gain implant similar to LGADs, but the segmentation into "pixels" is provided by AC-coupled electrodes, to enable high spatial resolution of hit via charge sharing. This allows for precise position resolution providing a 100\% fill factor and requiring fewer readout channels than traditional pixel sensors. The key to controlling the charge sharing is a precisely tuned n+ resistive layer as shown in figure~\ref{fig:RSD-LGAD-comp}. This differs from a Standard LGAD which uses a much more conductive, highly doped, n++ layer~\cite{RSD1Marta, NicoloBook}.

In several applications, such as adding RSD in a future HL-LHC tracker or tracking for a future collider, the RSD sensors will need to be able to operate in high fluence environments. The second RSD production at FBK \cite{RSDsecondProd} focused on preparing functional RSD sensors by optimizing readout pad shapes and improving the n+ resistive layer. However, these RSD sensors have not been studied under irradiation nor optimized for radiation tolerance. Here, we present a detailed irradiation study of this production of RSD sensors.

\section{Irradiation Campaign}

The second RSD production by FBK consisted of RSD sensors (with different readout shapes and pitches) and various test structures. In this campaign, two types of RSD sensors (450~\si{\micro\meter} and 1.3~\si{\milli\meter} pitch) and test structures were irradiated with protons and neutrons to three different fluences (four for test structures). Neutron irradiation was performed at the JSI TRIGA reactor and proton irradiation was performed using 23 MeV protons at the KIT ZAK cyclotron. The fluence points are shown in table~\ref{tab:irr_campaign} and were selected based on the expected fluences in the outer pixel layers inside of the Phase-2 CMS experiment during the HL-LHC.

\begin{table}[htbp]
    \centering
    \caption{RSD sensors were irradiated to the following nominal neutron fluences at the JSI TRIGA reactor and the measured 23 MeV proton fluences at the KIT ZAK synchotron.}
    \label{tab:irr_campaign}
    \begin{tabular}{c|c}
        \hline
        Nominal Neutron & Measured Proton \\
        Fluence [$10^{15}$~\si{\centi\meter^{-2}}] & Fluence [$10^{15}$~\si{\centi\meter^{-2}}] \\
        \hline
        1.0 & 0.6 \\
        2.0 & 1.0 \\
        3.5 & 1.8 \\ 
        \hline
        5.0 & 2.8 \\
        (test structures only) & (test structures only) \\
        \hline
    \end{tabular}
\end{table}

The sensors in this campaign come from four different wafers (W3, W4, W6, W14). All four wafers have shallow gain implants~\cite{NicoloBook}. W3 and W4 have a float zone substrate whereas W6 and W14 have an epitaxial substrate. The n+ layer doping increases among wafers in the following way: W3, W4 $<$ W6 $<$ W14.  Due to the similarities among wafers 3 and 4, sensors from these two wafers will be referred to as W3+4. Detailed descriptions of the wafer differences are available in \cite{LucaThesis}.

\section{Electrical Characterization}

\begin{figure}[htbp]
    \centering
    \includegraphics[width=0.7\linewidth]{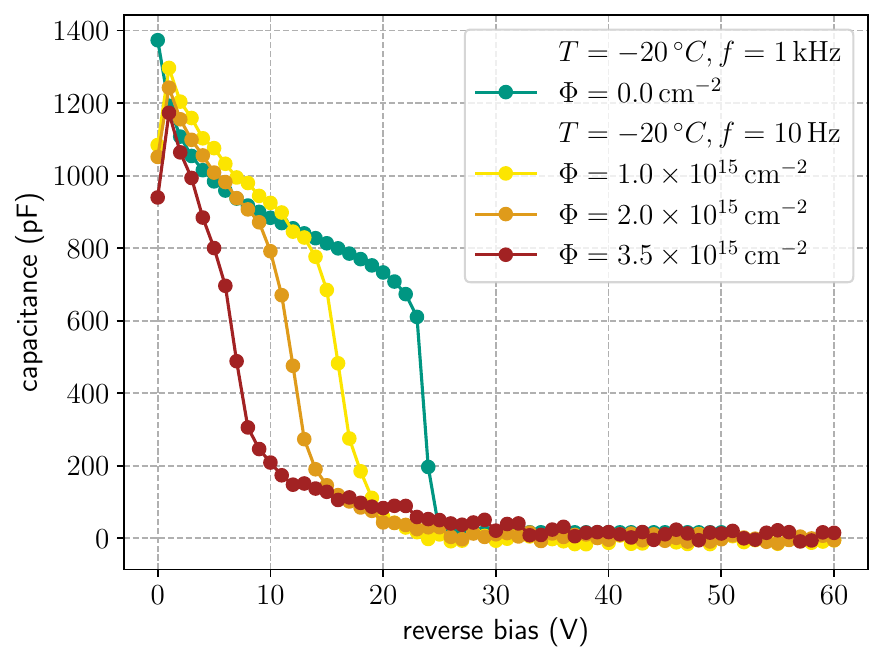}
    \caption{CV measurements for one wafer under neutron irradiation.}
    \label{fig:cv-meas}
\end{figure}

Before and after irradiation, the RSD sensors were characterized using the custom probe station at KIT to measure IV and CV curves. Dry air continuously flows inside of the probe station and all measurements took place with relative humidity $<5\%$ and a temperature of $-20$\si{\celsius}. CV measurements were performed at 1~\si{\kilo\hertz} before irradiation and 10~\si{\hertz} after irradiation to allow charges to untrap. An example of the CV measurements for one wafer, under neutron irradiation, is shown in figure~\ref{fig:cv-meas}

\subsection{Estimating the Fraction of Gain Layer Active Doping}

\begin{figure}[htbp]
    \centering
    \includegraphics[width=0.8\linewidth]{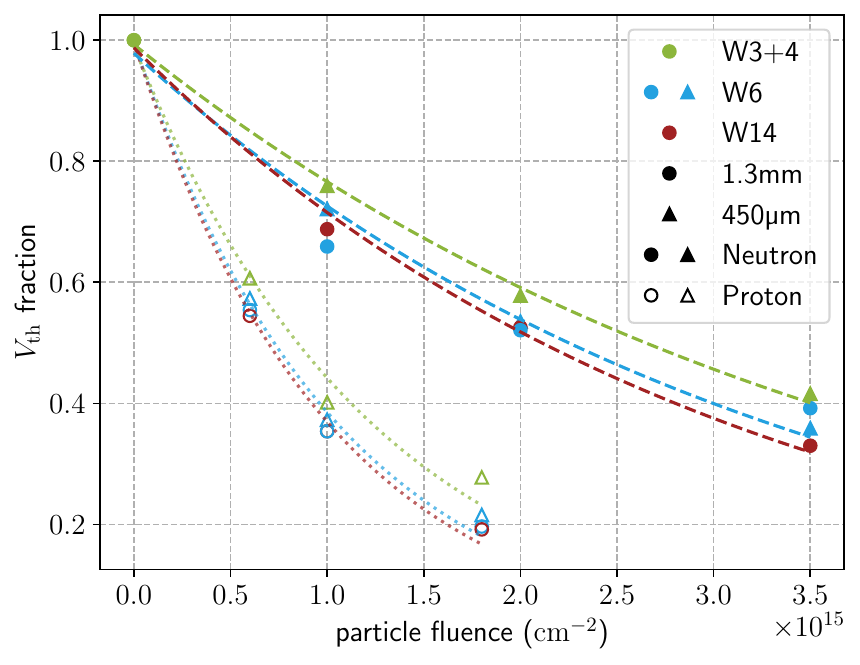}
    \caption{An estimate of the active fraction of gain layer, $V_{\textrm{th}}/V_{\textrm{GL}}$, versus nominal fluence for neutrons and measured fluence for protons. The results for each wafer are fit with an exponential. Figure based on \cite{MatteoLGAD}.}
    \label{fig:activeGainFrac}
\end{figure}

As the sensor depletes through the gain implant, the capacitance will drop significantly. By choosing a consistent capacitance value within this region, the variation in depletion voltage of the gain layer can be studied \cite{NicoloBook}. The capacitance value 300~\si{\pico\farad} was chosen as a proxy for estimating the gain layer depletion voltage variation. The voltage value corresponding to
300~\si{\pico\farad} was recorded as a $V_{\textrm{th}}$, with the value for unirradiated sensors being $V_{\textrm{GL}}=25$~\si{\volt}. The variation in depletion voltage can be observed with the fraction $V_{\textrm{th}}/V_{\textrm{GL}}$---as shown versus fluence in figure~\ref{fig:activeGainFrac}.

In order to measure how the gain deteriorates with increasing fluence, the results for each wafer were fit with an exponential function
\begin{equation}
    f(\Phi) = Be^{-c\Phi},
    \label{eq:expofit}
\end{equation}
where $B$ and $c$ are fit parameters and $\Phi$ is the particle fluence. The fit parameter $c$ then estimates the gain decrease. The $c$ coefficient values obtained from these fits are listed in table~\ref{tab:c_gain}. 

\begin{table}[htbp]
    \centering
    \caption{$c$ values measured in the exponential fits per wafer.}
    \begin{tabular}{c|c|c|c}
        \hline
        \multicolumn{4}{c}{\textbf{$c$ coefficient values} [$10^{-16}$~\si{\centi\meter^{2}}] } \\
        \hline
        Irradiation & W3+4 & W6 & W14 \\
        \hline
        Neutron & $2.6 \pm 0.1$ & $3.0 \pm 0.3$ & $3.2 \pm 0.2$ \\
        Proton & $8.1 \pm 0.7$ & $9.5 \pm 0.5$ & $9.9 \pm 0.4$ \\
        \hline 
    \end{tabular}
    \label{tab:c_gain}
\end{table}

\subsection{Comparison with LGAD Acceptor Removal Coefficients}

In a standard LGADs, this measurement technique is commonly used and $c$ is the acceptor removal coefficient \cite{NicoloBook}. Acceptor removal coefficients for LGADs with similar gain implants to the RSD are listed in \cite{NicoloBook} and are typically $\approx 5\times10^{-16}$~\si{\centi\meter^{2}} for neutron irradiation and $\approx 12\times10^{-16}$~\si{\centi\meter^{2}} for 23 MeV proton irradiation. The $c$ coefficient shown in table~\ref{tab:c_gain}, when compared to the typical values, would seem to indicate a gain layer that is more radiation-resistant than expected. Thus, leading to the question: why are the observed $c$ values for RSD lower than for standard LGADs with similar gain layers?

It is important to note that the value $c$ is extracted from CV measurements, which follow the evolution of sensor depletion---not just the gain layer. The depletion evolves from the p-n junction created from the n+ resistive layer and the p bulk. 

\section{Working Hypothesis}

\begin{figure}[htbp]
    \centering
    \includegraphics[width=0.7\linewidth]{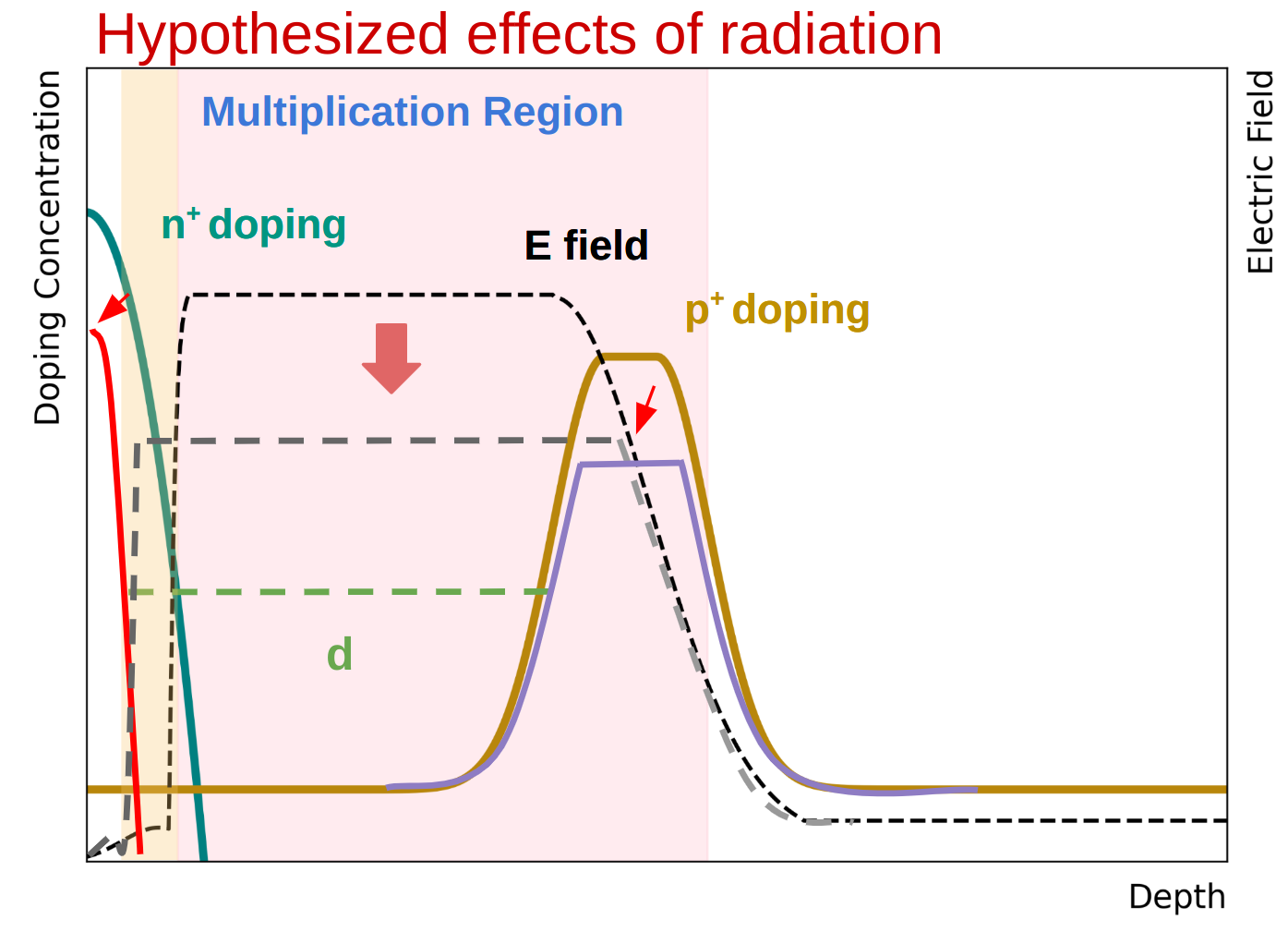}
    \caption{The hypothesized effects of irradiation on the RSD. The n+ layer, which has less doping than a standard LGAD, undergoes donor removal causing a wider multiplication region than that of a standard, irradiated LGAD.}
    \label{fig:hypothesis}
\end{figure}

Since the RSD features an n+ layer with a lower doping concentration compared to a standard LGAD, donor removal may influence the depletion voltage of the gain layer. The gain layer depletion voltage can be expressed with the following expression \cite{NicoloBook}

\begin{equation}
    V_{\textrm{GL}}\propto(1+2\frac{d}{w})N_{A}w^{2},
    \label{eq:gain_depletion}
\end{equation}
where $N_{A}$ is the gain layer active doping concentration, $w$ is the width of the gain implant, and $d$ is the depth of the gain implant relative to the n+ layer. After irradiation, the n+ layer shrinks, leading to a larger $d$ and a wider multiplication region with respect to an irradiated, standard LGAD. This effect leads to higher $V_{GL}$ values, explaining the smaller observed $c$ value. This hypothesis is shown visually in figure~\ref{fig:hypothesis}.

\section{Donor Removal}

Like acceptor states inside of the p+ gain implant, donor states in the n+ resistive layer are removed as the sensor is irradiated. This removal can be directly quantified by measuring the change in resistivity of the n+ layer with increasing fluence. During the campaign, `Van der Pauw' test structures for each wafer were irradiated. 

The `Van der Pauw` structures are directly connected to the n+ layer and allow for a four-point sheet resistance measurement.

\begin{figure}[htbp]
    \centering
    \begin{subfigure}{0.75\linewidth}
        \centering
        \includegraphics[width=\linewidth]{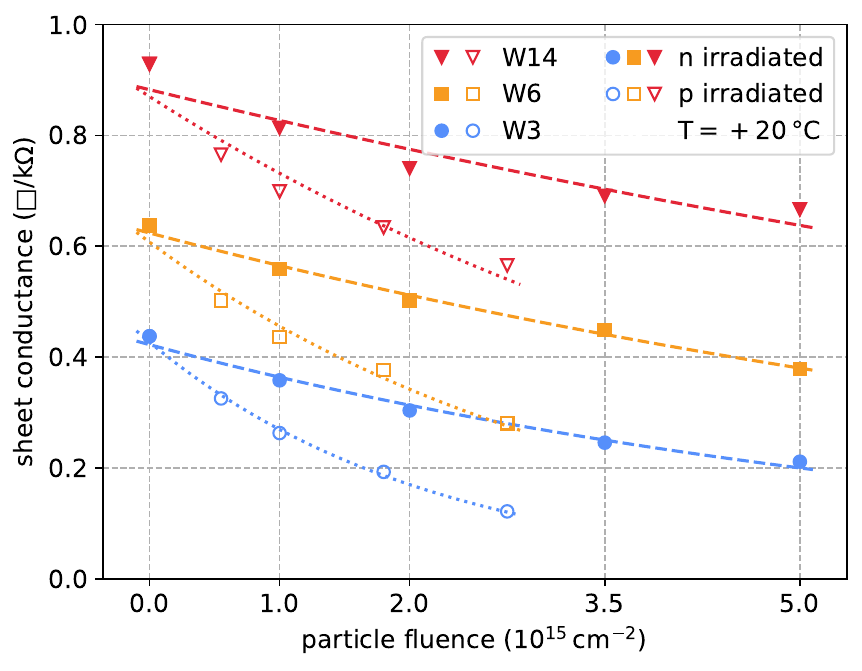}
    \end{subfigure}
    \caption{Sheet conductance versus fluence fit with an exponential curve in order to extract the donor removal coefficient $c_{\textrm{D}}$.}
    \label{fig:KIT_donor}
\end{figure}

The voltage drop in each structure was measured as the current was adjusted over a range of $-200$~\si{\micro\ampere} to $200$~\si{\micro\ampere} in $5$~\si{\micro\ampere} steps. The slope of the resulting line in voltage versus current is the sheet resistance ($R_{s}$). 
Donor removal can then be studied by plotting the sheet conductance ($1/R_{s}$) versus fluence, as shown in figure~\ref{fig:KIT_donor}. These measurements were performed both at KIT and at Perugia to eliminate possible biases from the measurement equipment.

Each wafer can then be fit with the exponential from equation~\ref{eq:expofit}, where the fit parameter $c$ becomes the donor removal coefficient $c_{D}$. The values of $c_{D}$ for each wafer are shown in table~\ref{tab:donor_removal}.

\begin{table}[htbp]
    \centering
    \caption{Donor removal coefficients $c_{D}$ measured in the exponential fits per wafer. Measurements of the test structures were performed at KIT and Perugia to ensure minimal bias form the measurement setup.}
    \begin{tabular}{c|c|c|c|c}
        \hline
        \multicolumn{5}{c}{\textbf{$c_{D}$ donor removal coefficient values} [$10^{-16}$~\si{\centi\meter^{2}}] } \\
        \hline
        Location & Irradiation & W3 & W6 & W14 \\
        \hline
        \multirow{2}{*}{KIT}
        & Neutron & $1.5\pm0.2$ & $1.0\pm0.2$ & $0.6\pm0.2$ \\
        & Proton & $4.6\pm0.4$ & $2.9\pm0.4$ & $1.7\pm0.3$ \\
        \hline 
        \multirow{2}{*}{Perugia}
        & Neutron & $1.6\pm0.2$ & $1.0\pm0.2$ & $0.7\pm0.1$ \\
        & Proton & $4.7\pm0.6$ & $2.9\pm0.4$ & $1.7\pm0.3$ \\
        \hline 
    \end{tabular}
    \label{tab:donor_removal}
\end{table}

The values in table~\ref{tab:donor_removal} indicate that donor removal is higher in less doped n+ layers. This result mirrors what has been observed for acceptor removal, where the $c$ coefficient is also higher for lower acceptor densities ~\cite{NicoloBook}.

\section{Charge Sharing Studies with the Transient Current Technique (TCT)}

The Transient Current Technique (TCT) was used to further study the effects of irradiation on the RSD sensors. Of particular importance is to study the effect on charge sharing between the AC readout pads. RSD sensors rely on charge sharing to precisely reconstruct the position of a particle---as an ionizing particle passes through the sensor, electrons will follow the path of least resistance to induce a signal on the readout pads \cite{SivieroPosition, RSD1Marta}. Precise reconstruction can be done in a number of ways described in \cite{SivieroPosition}, but dramatic changes after irradiation would require constant recalibration of the sensors. Signal charge sharing can be studied using the TCT; in this case, an infrared laser of 1055~\si{\nano\meter} is used to create electron-hole pairs in the sensor and observe the resulting signal on the readout pads. These studies are still ongoing and these are only early results. 

\begin{figure}[htbp]
    \centering
    \includegraphics[width=0.75\linewidth]{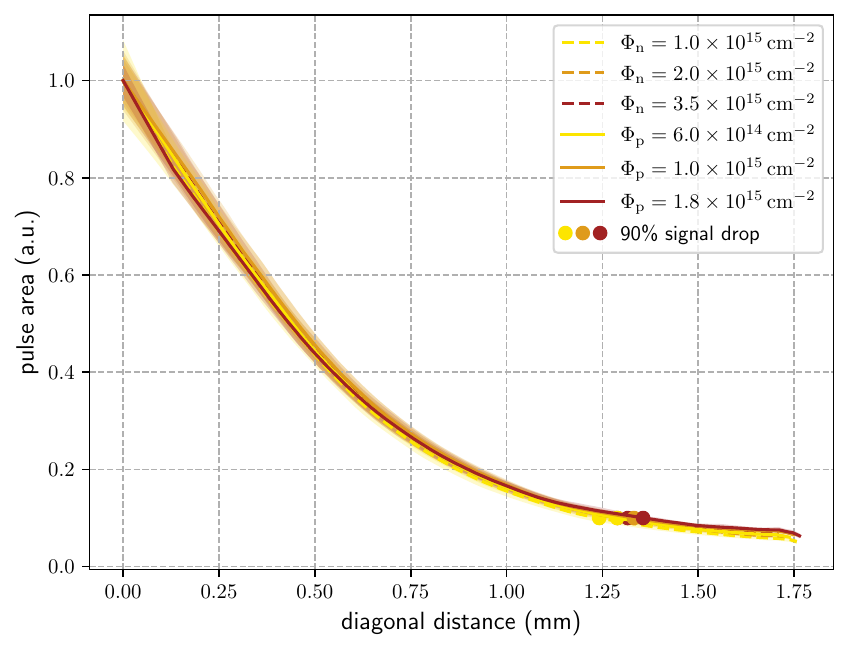}
    \caption{The signal area as the TCT laser moves diagonally outward from the center of an AC readout pad. The distributions for different fluences are normalized to compare the shapes and the point where the signal decreases by 90\% of the original is marked with a point. This is currently only shown for W14.}
    \label{fig:rad-signal}
\end{figure}

The first study focuses on how the charges move in the sensor by observing the signal (area under the pulse) in one AC pad as the laser moves diagonally outward. The shapes of the resulting distributions are compared in figure~\ref{fig:rad-signal}. The point at which the signal area decreases by 90\% is marked with a point. The normalized distributions and 90\% points are remarkably similar, yielding an excellent result---the charge sharing appears to be the same. This study was only performed for W14 so far and must be expanded to more wafers and additional studies on charge spreading are planned for the next months.

\section{Conclusions}

In this study, several parameters of irradiated RSD sensors were measured.

First, the gain layer depletion voltage was measured for proton and neutron irradiated RSD sensors, using CV measurements. Two important trends were observed: (i) the coefficients $c$, which estimate $V_{GL}$ decrease with irradiation, were found to be lower than those measured in LGADs with similar gain implants (the acceptor removal coefficients), (ii) the coefficients $c$ were found to be lower for RSD sensors with a more resistive n+ layer. Both results are attributed to the effect of donor removal in the n+ layer, a process that is negligible in standard LGADs. As with LGADs, proton radiation is more damaging than neutron radiation in RSD sensors.

Then, the donor removal was studied using Van der Pauw structures at several irradiation levels and initial donor densities. The results clearly show that lower donor densities have a higher coefficient $c$, a trend similar to acceptor removal. A higher donor removal coefficient for less doped n+ layer accounts for the observed variations in the depletion voltage of the gain layer:  the decrease in $V_{GL}$ is less pronounced as the depleted gap in the gain layer becomes wider with irradiation, as shown in equation~\ref{eq:gain_depletion}. 

Finally, charge sharing in new and irradiated RSD sensors was studied using a TCT setup. Our preliminary results indicate that irradiation does not alter the sharing mechanism.

\section*{Acknowledgments}

We would like to acknowledge the support of the following funding agencies and collaborations: European Union’s Horizon Europe Research and Innovation programme Grant Agreement No 101057511 (EURO-LABS); Alexander von Humboldt Stiftung; KIT KCETA Ausschreibung Sachmittel; INFN - CSN5 RSD project; Dipartimenti di Eccellenza, Univ. of Torino (ex L. 232/2016, art. 1, cc. 314, 337); European Union -Next Generation EU, Mission 4 component 2, CUP C53D23001510006



\end{document}